\documentclass[prb,twocolumn,superscriptaddress]{revtex4}
\usepackage{color}
\usepackage{amsfonts}
\usepackage{amsmath}
\usepackage{amssymb}
\usepackage{graphicx}

\begin{document}

\title{Conductance scaling of junctions of Luttinger-liquid wires out of
equilibrium}
\author{D.~N. Aristov}
\affiliation{NRC ``Kurchatov Institute", Petersburg Nuclear Physics Institute, Gatchina
188300, Russia}
\affiliation{Institute for Nanotechnology, Karlsruhe Institute of Technology, 76021
Karlsruhe, Germany }
\affiliation{St.Petersburg State University, 7/9 Universitetskaya nab., 199034
St.~Petersburg, Russia}
\author{P. W\"olfle}
\affiliation{Institute for Nanotechnology, Karlsruhe Institute of Technology, 76021
Karlsruhe, Germany }
\affiliation{Institute for Condensed Matter Theory, Karlsruhe Institute of Technology,
76128 Karlsruhe, Germany}

\begin{abstract}
We develop the renormalization group theory of the conductances of N-lead
junctions of spinless Luttinger-liquid wires as functions of bias voltages
applied to N independent Fermi-liquid reservoirs. Based on the perturbative
results up to second order in the interaction we demonstrate that the
conductances obey scaling. The corresponding renormalization group $\beta -$%
functions are derived up to second order.\ 
\end{abstract}

\maketitle

\section{Introduction}

\label{intro}

The charge transport through junctions connecting quantum wires modeled by
the Tomonaga-Luttinger liquid model (TLL) has been studied intensely over
the past several decades. In the linear response regime it has been shown
since the first studies of a two-lead junction \cite%
{Kane1992,Safi1995,Furusaki1996,Sassetti1996} that the conductance obeys
scaling as a function of temperature, at least in the vicinity of certain
special values of the conductance. This behavior is captured in the
framework of a renormalization group (RG) formulation, where the special
values are identified as fixed points of the RG flow. Initially the flow
equations were derived within the bosonization approach. The latter has the  
{difficulty} that the conductance of a wire of finite length depends on the
contact resistances at the links to the external charge reservoirs
(accounting for the transition of bosonic excitations into fermionic
quasiparticles), which so far have not been determined. Alternatively, a
purely fermionic formulation may be used, which avoids the problem of
contact resistance.

The latter approach has been pioneered by \cite{Yue1994} in the limit of
weak interaction and was later extended to arbitrary coupling strength by 
\cite{Aristov2009}. In simple words, the standard procedure for how to
derive the RG flow equations, e.g. for the two-lead junction, is to first
calculate the conductance $G_{0}(\theta )$ in the absence of interaction, as
a function of the parameter $\theta $ (or of several parameters in the case
of multi-lead junctions) determining the scattering strength of the
junction. Then the conductance is calculated in perturbation theory in the
interaction, allowing to identify the linear logarithmic corrections (for
example $\propto \ln (\omega _{0}/T)$, if the temperature $T$ is cutting off
the infrared singularities of the theory and $\omega _{0}$ is an ultraviolet
cutoff), and extract the RG $\beta -$function as $\beta =-dG/d\ln T$ at $\ln
(\omega _{0}/T)=0$. The resulting function $\beta $ depends on the
interaction strength $\alpha $ and on the parameter $\theta $ characterizing
the junction. Inverting the functional dependence $G_{0}(\theta )$, the
parameter $\theta $ may be expressed by $G_{0}$, which in the sense of the
RG structure may be replaced by the renormalized $G$ at scale $T$. This
procedure may be justified within a more rigorous scheme using ideas first
formulated by \cite{Callan1970,Symanzik1970}, see \cite{Aristov2009}. In
order to explicitly demonstrate the scaling property, it should be shown
that all terms of powers higher than linear in $\ln (\omega _{0}/T)$,
appearing in perturbation theory, are generated by iteration of the RG
equations. For the case of the conductance of a two-lead junction this has
been verified up to order $(\ln (\omega _{0}/T))^{3}$ \cite{Aristov2009}.
The approach sketched above has also been applied to multilead junctions,
such as the Y-junctions in the weak coupling limit \cite%
{Lal2002,Aristov2010,Shi2016} and at strong coupling \cite{Aristov2011a}, and
chiral Y-junctions \cite{Aristov2012, Aristov2013,Aristov2017}, as well as X-junctions%
\cite{Aristov2016a}.

The results on transport through Y-junctions obtained in our previous work are generally in good agreement with results obtained by the bosonization method (BM) in the linear response regime 
\cite{Nayak1999, Chamon2003, Oshikawa2006,Das2011}, keeping in mind that the overall magnitude of the current can not be determined accurately by BM, as mentioned above. In the few cases where discrepancies have arisen, such as in the limit of very strong attractive interaction, \cite{Aristov2011a} the BM calculation employed additional assumptions which we believe to be incorrect. There are only few works on transport through Y-junctions out of equilibrium, in which scaling has been assumed to exist without proof  \cite{Safi2009,Aristov2017a}. 

The fermionic transport formulation is general and physically appealing. It
may be extended to systems out of equilibrium in a natural way. Transport
through a two-lead junction at finite voltage bias $V$ and low temperatures
has been considered in \cite{Aristov2014}, with results in agreement with
those of other methods, where applicable \cite{Egger2000,Dolcini2003,
Dolcini2005, Metzner2012}. Not too surprisingly, one finds the conductance
following a power law in $V$, with exponent identical to the one governing
the temperature power law of the linear response conductance. Recently,
transport through a Y-junction out of equilibrium has been studied by
assuming the scaling property to hold in this case as well\cite{Aristov2017a}%
. The latter assumption is not trivial, if only for the following reason:
the Y-junction is connected to three charge reservoirs held at three
different chemical potentials, in general. Consequently, the two independent
conductances depend on two independent bias voltages $V_{a},V_{b}$ . There
appears to be scaling of the conductances in both variables, $V_{a}$ and $%
V_{b}$. {The question is then how the scaling in $T$ is expressed in terms
of voltage, i.e. which of the two voltages or both enter the corresponding
formulas. } Indeed it has been found in \cite{Aristov2017a} that the scaling
as a function of $V_{a},V_{b}$ may not be obtained from the scaling of the
linear response conductance with $T$ by any simple recipe.

In the present paper we demonstrate that scaling in the case of multi-lead
junctions out of equilibrium is valid, by explicitly calculating the terms
of second power in the scaling variable $\Lambda =\ln (\omega _{0}/\epsilon )
$ in the conductances of a symmetric Y-junction in second order in the
interaction. Here $\omega _{0}$ and $\epsilon $ are ultraviolet and infrared
cutoffs in energy. We then show that all of these terms are generated by the
RG equations, proving the validity of scaling, at least up to this order.

\section{The Model}

We consider a system of spinless fermions in one dimension, interacting in
each of $N$ quantum wires in the region $a<|x_{j}|<L$, $j=1,..,N$\ ,
adiabatically connected to $N$ charge reservoirs at $|x_{j}|>L$. The $N$
wires are connected by a junction located in the narrow regime $|x_{j}|<a$,
which scatters the fermions as described by an $S$ matrix with elements $%
S_{jk}$, where $j,k=1,...N$\ .

We assume the interaction to be described by a TLL model in the form 
\begin{equation}
H=\int_{0}^{\infty }dx\sum_{j=1}^{N}[H_{j}^{0}(x)+H_{j}^{int}(x)\Theta
(x;a,L)]
\end{equation}%
where 
\begin{equation*}
H_{j}^{0}(x)=v_{j}[\psi _{j,in}^{\dag }(x)i\nabla \psi _{j,in}(x)-\psi
_{j,out}^{\dag }(x)i\nabla \psi _{j,out}(x)]
\end{equation*}%
and 
\begin{equation*}
H_{j}^{int}(x)=2\pi v_{j}\alpha _{j}[\psi _{j,in}^{\dag }(x)\psi
_{j,in}(x)\psi _{j,out}^{\dag }(x)\psi _{j,out}(x)]
\end{equation*}%
Here $v_{j}$ is the Fermi velocity, $\alpha _{j}$ is the interaction
constant in lead $j$ and $\Theta (x;a,L)=1$ in the interval $a<|x|<L$ and
zero elsewhere. The fermionic field operators $\psi _{j,\eta _{j}}^{\dag }(x)
$ create particles at position $x$ in scattering states $|j,\eta _{j};\omega
\rangle $ of energy $\omega $, in wire $j$ and with chirality $\eta _{j}=\pm
1$, labeling incoming ($\eta _{j}=-1$) and outgoing ($\eta _{j}=+1$) states.
In the following we will put $v_{j}=1$ for simplicity. The outgoing fermion
operators are connected with the incoming ones by the $S$ matrix, $\psi
_{j,out}(0)=S_{jk}\psi _{k,in}(0)$.

The perturbation theory will be formulated in the language of Keldysh matrix
single particle Green's functions

\begin{equation}
\underline{G}=\left( 
\begin{array}{cc}
G^{R} & G^{K} \\ 
0 & G^{A}%
\end{array}%
\right)
\end{equation}%
in the non-interacting limit. The Green's functions depend on energy $\omega 
$, position $x$, wire index $j$ and chirality $\eta _{j}$, $\underline{G}=%
\underline{G}_{\omega }(l_{\eta _{l}},y;j_{\eta _{j}},x)$.

The Green's functions for each pair of chiralities $\eta _{l},\eta _{j}$ are
given by

\begin{equation}
\begin{aligned} \underline{G}_{\omega }(l_{+},y;j_{+},x) & =-ie^{i\omega
(y-x)} \begin{bmatrix} \theta (y-x)\delta _{lj} , & S_{lm}S_{jm}^{\ast
}h_{m} \\ 0, & -\theta (x-y)\delta _{lj}\end{bmatrix}, \\
\underline{G}_{\omega }(l_{+},y;j_{-},x) & =-ie^{i\omega (y+x)}
\begin{bmatrix} S_{lj} ,& S_{lj}h_{j} \\ 0 ,& 0\end{bmatrix}, \\
\underline{G}_{\omega }(l_{-},y;j_{+},x) & =-ie^{-i\omega (y+x)}
\begin{bmatrix} 0,& S_{jl}^{\ast }h_{l} \\ 0, & - S_{jl}^{\ast }\end{bmatrix}, \\ 
\underline{G}_{\omega
}(l_{-},y;j_{-},x)& = - ie^{i\omega (x-y)} 
\delta_{lj} 
\begin{bmatrix}  \theta (x-y), &  h_{l} \\ 0 ,&- \theta (y-x) \end{bmatrix},
\end{aligned}  \label{def:Green}
\end{equation}
%%%%%%% 
where $h_{l}(\omega )=\tanh [(\omega -\mu _{l})/2T]$ is the Keldysh
function, with $\mu _{l}$ the chemical potential in wire $l$ ;  summation
over $m$ is implied in the first line of \eqref{def:Green}.  The functions $%
h_{l}(\omega )$ carry the information on the out of equilibrium conditions.

\section{Derivation of Renormalization Group equations for the conductances}

We consider the conductances $G_{j},$ $j=1,..N_{G}$ of a $N-$lead junction
out of equilibrium (here $N_{G}<N^{2}$ is the number of independent
conductances).\ In order to show that the conductances obey scaling, we
follow the reasoning first developed by Callan and Symanczik Ref. \cite%
{Callan1970,Symanzik1970} and apply it to the problem at hand. Assume that
we know the perturbation series of the conductances $G_{j}$. The
conductances depend on the scattering properties of the junction, expressed
by the S-matrix elements $S_{lj}$ (which may be expressed in terms of the
conductances $G_{j}^{0}$ of the system in the absence of interactions), and
on the interaction (coupling constant $\alpha $). The possible existence of
scaling is signaled by the appearence of powers of the scaling variable $%
\Lambda =\ln (\omega _{0}/\epsilon )$. The perturbation series of $G_{j}$ in
terms of the interaction has the general form%
\begin{equation}
G_{j}=G_{j}^{0}+\alpha A_{j}(\{G_{i}^{0}\})+\alpha
^{2}B_{j}(\{G_{i}^{0}\})+O(\alpha ^{3})  \label{G-pert}
\end{equation}%
where $A_{j},B_{j}$ are polynomials of first and second order in the scaling
variable $\Lambda $, respectively, polynomials of the bare conductances $%
G_{j}^{0}$ , functions of the $N_{V}$ independent bias voltages $V_{j}$, and
of additional coupling constants $\alpha _{j}$\ in the form of ratios $%
\alpha _{j}/\alpha $. Now we invert the series to express the conductances $%
G_{j}^{0}$ in the non-interacting limit in terms of the full conductances$\
G_{j}$

\begin{equation}
G_{j}^{0}=G_{j}+\alpha \overline{A}_{j}(\{G_{i}\})+\alpha ^{2}\overline{B}%
_{j}(\{G_{i}\})+O(\alpha ^{3})  \label{G0-pert}
\end{equation}
where $\overline{A}_{j},\overline{B}_{j}$ are again polynomials of first and
second order in $\Lambda $.\ By substituting $G_{j}^{0}$ as given in Eq.\ (%
\ref{G0-pert})\ into Eq.\ (\ref{G-pert})\ we find the following relations

\begin{equation}
\begin{aligned} \overline{A}_{j}(\{G_{i}\}) &=-A_{j}(\{G_{i}\}) \\
\overline{B}_{j}(\{G_{i}\}) &=-B_{j}(\{G_{i}\})-\sum_{l}\frac{\partial
A_{j}(\{G_{i}\})}{\partial G_{l}}\overline{A}_{l}(\{G_{i}\}) \end{aligned}
\end{equation}

We now use that $G_{j}^{0}(\{G_{i}\})$ must be independent of $\Lambda $

\begin{equation}
\frac{dG_{j}^{0}(\{G_{i}\})}{d\Lambda }=0=\frac{\partial G_{j}^{0}(\{G_{i}\})%
}{\partial \Lambda }+\sum_{l}\frac{\partial G_{j}^{0}(\{G_{i}\})}{\partial
G_{l}}\frac{\partial G_{l}}{\partial \Lambda }
\end{equation}%
\ to find the renormalization group equation

\begin{equation}
\frac{\partial G_{j}}{\partial \Lambda }=-\sum_{l}\left( \frac{\partial
G_{j}^{0}(\{G_{i}\})}{\partial G_{l}}\right) ^{-1}\frac{\partial
G_{l}^{0}(\{G_{i}\})}{\partial \Lambda }
\end{equation}%
The inversion of the matrix $D_{jl}=\partial G_{j}^{0}(\{G_{i}\})/\partial
G_{l}$ is obtained from Eq.\ (\ref{G0-pert})\ as

\begin{equation}
\left( \frac{\partial G_{j}^{0}(\{G_{i}\})}{\partial G_{l}}\right)
^{-1}=\delta _{jl}+\alpha \frac{\partial A_{j}(\{G_{i}\})}{\partial G_{l}}
\end{equation}%
\ Substituting $\partial G_{l}^{0}(\{G_{i}\})/\partial \Lambda $ we finally
get

\begin{equation}
\begin{aligned} \frac{\partial G_{j}}{\partial \Lambda } &=\alpha
\frac{\partial A_{j}(\{G_{i}\})}{\partial \Lambda } \\ &+\alpha ^{2}\left(
\frac{\partial B_{j}(\{G_{i}\})}{\partial \Lambda }-\sum_{l}\frac{\partial
^{2}A_{j}(\{G_{i}\})}{\partial \Lambda \partial
G_{l}}A_{l}(\{G_{i}\})\right) \end{aligned}  \label{dG/dLambda}
\end{equation}%
In order for scaling to hold, the r.h.s. of Eq.\ (\ref{dG/dLambda}) should
not depend on $\Lambda $. More specifically, it should be $\mathcal{O}%
(\Lambda ^{0})$, possibly containing Heaviside step functions (see below),
indicating a stop of the RG flow. Verifying this property amounts to proving
the validity of scaling up to the order considered.

\section{Symmetric Y-junction out of equilibrium}

We now apply the above general derivation of RG equations for the
conductances to a concrete example.\ We consider charge transport through a
Y-junction with a symmetric main wire (labelled $1,2$) contacted by a tip
wire at the center ($3$) (see Fig.1). The three half-wires of length $L$ are
adiabatically connected with reservoirs kept at chemical potentials $\mu _{j}
$, $j=1,2,3$. We assume that there is no interaction within the junction of
radius $a$. The scattering states of each wire are labeled by wire index $j$%
, chirality $\eta _{j}=+,-$ (outgoing or ingoing), energy $\omega $, and
position $x>0$ in the interval $[a,L]$. The junction is symmetric in the
sense that the interaction constants $\alpha _{1}$ and $\alpha _{2}$ in arms 
$1$ and $2$, respectively, are equal to each other, $\alpha _{1}=\alpha
_{2}\equiv \alpha $. The third arm of the junction is a tunneling-tip wire,
with interaction constant $\alpha _{3}$, which we will assume to vanish in
the following. We define currents $J_{j}$ flowing from the reservoirs toward
the junction.

\begin{figure}[tbp]
\includegraphics[width=0.85\columnwidth]{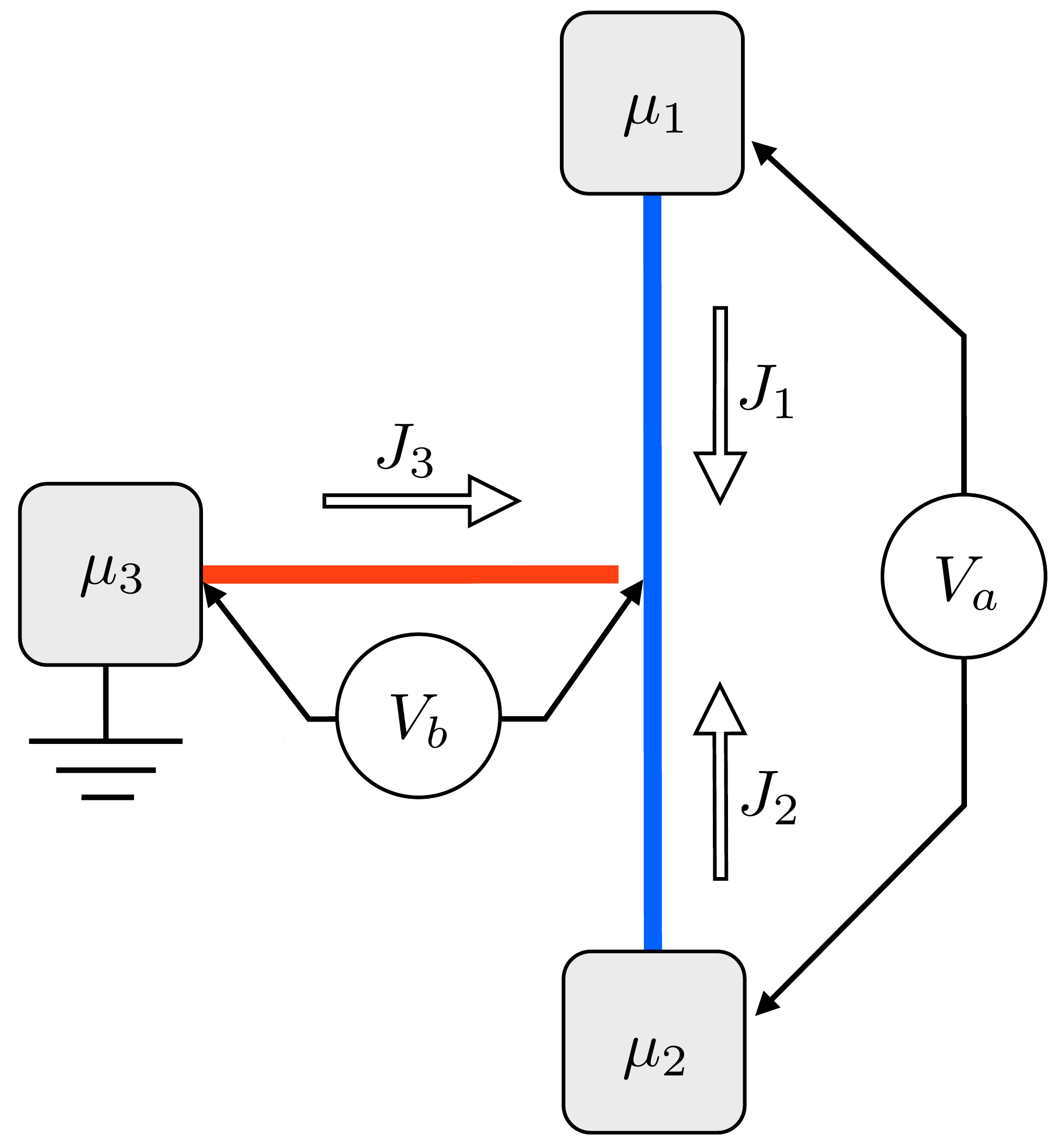}
\caption{Setup of the Y-junction out of equilibrium. The main wire is shown
as a blue vertical line, the tunneling tip as a red horizontal line. The
reservoirs at the chemical potentials $\protect\mu _{a,b,c}$ are depicted as
gray blocks, with the currents $J_{a,b,c}$ flowing out from them in the
presence of the bias voltages $V_{a}$ and $V_{b}$. }
\end{figure}

The $S$ matrix of the symmetric Y-junction may be parametrized as follows 
\begin{equation}
S=%
\begin{pmatrix}
r_{1}, & t_{1}, & t_{2} \\ 
t_{1}, & r_{1}, & t_{2} \\ 
t_{2}, & t_{2}, & r_{2}%
\end{pmatrix}%
\,.  \label{Smatrix}
\end{equation}
The symmetric form of interaction, $\alpha_{1}=\alpha_{2}$, keeps the
renormalized S matrix in symmetric form \eqref{Smatrix}. We use the
parametrization

\begin{equation}
\begin{aligned} r_{1}& = \tfrac{1}{2} (\cos \vartheta +e^{-i \psi } ) ,
\quad t_{1} = \tfrac{1}{2} (\cos \vartheta -e^{-i \psi } ) \,, \\ t_{2} & =
\tfrac{i }{\sqrt{2}} \sin \vartheta , \quad r_{2} = \cos \vartheta \,.
\end{aligned}  \label{paramSmat}
\end{equation}

It is convenient to introduce two independent currents $J_{a,b}$ and two
independent bias voltages $V_{a,b}$ as follows: 
\begin{equation}
J_{a}=\frac{1}{2}(J_{1}-J_{2}),\quad V_{a}=\mu _{1}-\mu _{2}
\end{equation}%
for the main wire and 
\begin{equation}
J_{b}=\frac{1}{3}(J_{1}+J_{2}-2J_{3})=-J_{3},\quad V_{b}=\frac{1}{2}(\mu
_{1}+\mu _{2})
\end{equation}%
for the tunneling tip. The conductances $G$ are then defined as 
\begin{equation}
J_{a}=G_{a}V_{a}+G_{ab}V_{b}, \quad J_{b}=G_{ba}V_{a}+G_{b}V_{b}
\end{equation}%
It is found that in the symmetric setup $G_{ab}$ and $G_{ba}$ appear due to
asymmetry produced by the voltages, they may be expressed in terms of the
diagonal conductances $G_{a},G_{b}$ and therefore do not flow independently.
We therefore do not consider the off-diagonal conductances in the following.
In terms of parametrization \eqref{paramSmat} the conductances are given by  
$G_{a} = (1-\cos\vartheta\cos\psi)/2$, $G_{b} = \sin^{2}\vartheta $.

\subsection{Perturbation theory results and RG-equation in first order}

As shown in \cite{Aristov2017a} the diagonal conductances in first order are
given by

\begin{equation}
\begin{aligned} G_{a}(\epsilon ) &=G_{a}^{0}+\alpha \,
(a_{1}^{0}\,\Lambda_{a} +a_{2}^{0} (\Lambda_{b+} + \Lambda_{b-})) \,, \\
G_{b}(\epsilon ) & =G_{b}^{0}+\alpha\, a_{3}^{0}\, (\Lambda_{b+} +
\Lambda_{b-}) \end{aligned}  \label{corr1stOrder}
\end{equation}
where $G_{a,b}^{0}=G_{a,b}(\omega _{0})$ and we use a shorthand notation $%
\Lambda_{a} = \ln( \omega_{0} / \max [\epsilon, V_{a} ])$, $\Lambda_{b\pm} =
\ln( \omega_{0} / \max [\epsilon, |V_{b\pm}|])$, with $V_{b\pm }=V_{b}\pm
V_{a}/2$ .  Here we defined $a_{j}^{0}=a_{j}(G_{a}^{0},G_{b}^{0})$ with  
\begin{equation}
\begin{aligned} a_{1}(G_{a},G_{b}) &=-2G_{a}(1-G_{a})+\tfrac{1}{2}G_{b}, \\
a_{2}(G_{a},G_{b}) &=-\tfrac{1}{8}[1-G_{a} + g_{3} (1-2G_{a}) ]G_{b}, \\
a_{3}(G_{a},G_{b}) &=-\tfrac{1}{2}[1-G_{a}-\tfrac12 G_{b} +
g_{3}(1-G_{b})]G_{b}. \end{aligned} 
\label {defs:a123}
\end{equation}
and $g_{3} = \alpha_{3}/\alpha$.

Differentiating \eqref{corr1stOrder} with respect to $\Lambda =\ln (\omega
_{0}/\epsilon )$ and replacing $a_{j}^{0} \to a_{j}$, we obtain the RG $%
\beta -$functions as  
\begin{equation}
\begin{aligned} \frac{\partial G_{a}}{\partial \Lambda } &=\beta
_{a}(G_{a},G_{b}) \\ &=\alpha [ a_{1}(G_{a},G_{b})\,\theta_{a}
(\epsilon)+a_{2}(G_{a},G_{b})\,\theta _{+}(\epsilon )]\, , \\ \frac{\partial
G_{b}}{\partial \Lambda } &=\beta _{b}(G_{a},G_{b})=\alpha\,
a_{3}(G_{a},G_{b})\,\theta _{+}(\epsilon )\, . \end{aligned}  \label{Beta}
\end{equation}
with $\theta _{a }(\epsilon )=\theta (\epsilon - V_{a})$ and 
\begin{equation*}
\theta _{+ }(\epsilon )=\theta (\epsilon -|V_{b-}|)+ \theta (\epsilon
-V_{b+}) \,,
\end{equation*}

The effect of the $\theta-$functions is to define different forms of the
functions $A_{a,b}$ in different intervals of $\epsilon $, and hence $%
\Lambda $. For example, for given $V_{a}>2V_{b}>0$ we have $\theta
_{+}(\epsilon )=2$ for $\epsilon >$ $V_{b}+V_{a}/2$ and $\theta_{a}
(\epsilon)=1$ for $\epsilon >$ $V_{a}$\ (interval $I$), $\epsilon >$ $%
V_{b}+V_{a}/2$ and $\theta_{a} (\epsilon)=0$ for $\epsilon <$ $V_{a}$
(interval $II$), $\theta _{+}(\epsilon )=1$ for $|V_{b}-V_{a}/2|<\epsilon
<V_{b}+V_{a}/2$ (interval $III$)\ and $\theta _{+}(\epsilon )=0$ for $%
|V_{b}-V_{a}/2|>\epsilon $ (interval $IV$).

If the differential equations \eqref{Beta} are valid, then the calculated
second order corrections, $\sim \alpha^{2 }\Lambda^{2}$, are cancelled in
the above procedure, leading to Eq.\ \eqref{dG/dLambda}. Alternatively, we
may compare these corrections with the predicted form stemming from equation %
\eqref{Beta}.

The conductances $G_{a,b}$ up to second order in $\Lambda $ are determined
by solving the RG-equations iteratively. We substitute the first order
results, Eq.\ \eqref{corr1stOrder}, into the $\beta -$functions and
integrate. We get 
\begin{equation*}
G_{a,b}=G_{a,b}^{0}+\alpha G_{a,b}^{\prime }+ \tfrac 12 \alpha ^{2}G_{a,b}^{\prime
\prime },
\end{equation*}%
where for $n=a,b$ 
\begin{equation}
\begin{aligned} G_{n}^{\prime } & = \int _{\omega_{0}}^{\epsilon}
\frac{d\epsilon'}{\epsilon'} \beta _{n}(G^{0}_{a},G^{0}_{b}) _{\epsilon'} \\
G_{n}^{\prime\prime } & = 2 \int _{\omega_{0}}^{\epsilon}
\frac{d\epsilon'}{\epsilon'} \int _{\omega_{0}}^{\epsilon'}
\frac{d\epsilon''}{\epsilon''} \Big [ \frac {\partial \beta
_{n}(G^{0}_{a},G^{0}_{b})_{\epsilon'} }{\partial G^{0}_{a} } \beta
_{a}(G^{0}_{a},G^{0}_{b}) _{\epsilon''} \\ & + \frac{ \partial \beta
_{n}(G^{0}_{a},G^{0}_{b})_{\epsilon'} }{ \partial G^{0}_{b}} \beta
_{b}(G^{0}_{a},G^{0}_{b})_{\epsilon''} \Big ] \end{aligned}  \label{iterat}
\end{equation}%
The expression in square brackets here reads 
\begin{equation}
\begin{aligned} G_{a}^{\prime \prime } &= \frac {\partial a_{1}^{0}
}{\partial G^{0}_{a}} a_{1}^{0} \,\theta_{a} (\epsilon'') \theta_{a}
(\epsilon') + \frac {\partial a_{2}^{0} }{\partial G^{0}_{a}} a_{1}^{0}
\theta _{+}(\epsilon' ) \theta_{a} (\epsilon'') \\ &+\left ( \frac {\partial
a_{1}^{0} }{\partial G^{0}_{a}} a_{2}^{0} + \frac {\partial a_{1}^{0}
}{\partial G^{0}_{b}} a_{3}^{0} \right) \, \theta _{a}(\epsilon' ) \theta
_{+}(\epsilon'' ) \\ & + \left ( \frac {\partial a_{2}^{0} }{\partial
G^{0}_{a}} a_{2}^{0} + \frac {\partial a_{2}^{0} }{\partial G^{0}_{b}}
a_{3}^{0} \right )\, \theta _{+}(\epsilon' ) \theta _{+}(\epsilon'' )
\end{aligned}  \label{iteratA}
\end{equation}%
and 
\begin{equation}
\begin{aligned} G_{b}^{\prime \prime } &= \frac {\partial a_{3}^{0}
}{\partial G^{0}_{a}} a_{1}^{0} \, \theta _{+}(\epsilon' )\theta_{a}
(\epsilon'') \\ & + \left( \frac {\partial a_{3}^{0} }{\partial G^{0}_{a}}
a_{2}^{0} + \frac {\partial a_{3}^{0} }{\partial G^{0}_{b}} a_{3}^{0}
\right) \theta _{+}(\epsilon' ) \theta _{+}(\epsilon'' ) \end{aligned}
\label{iteratB}
\end{equation}%
% Here we defined $a_{j}^{(1)}=a_{j}(G_{a}^{(1)},G_{b}^{(1)})$. 

Eqs. \eqref{iteratA}, \eqref{iteratB} contain $\theta $ functions in four
different combinations. One of them, $\theta _{a}(\epsilon ^{\prime })\theta
_{a}(\epsilon ^{\prime \prime })$ results simply in $\Lambda _{a}^{2}/2$.
Others, e.g. $\theta _{+}(\epsilon ^{\prime })\theta _{a}(\epsilon ^{\prime
\prime })$ lead to more complicated expressions and depend on the relation
between $V_{a}$, $V_{b}$. As was shown in \cite{Aristov2017a}, there are two
most interesting cases. In one of them, with $V_{a}\agt V_{b}$, we can let $%
\theta _{+}(\epsilon )\simeq 2\theta (\epsilon -V_{a})$, which results in
unique value of logarithm, $\Lambda _{a}$. In another regime, $V_{a}\ll V_{b}
$, we let $\theta _{+}(\epsilon )\simeq 2\theta (\epsilon -V_{b})$ and
obtain two values, $V_{a}$ and $V_{b}$.

In this second regime we can express the integrals appearing in %
\eqref{iterat} as 
\begin{equation}
\begin{aligned} \int d\Lambda' d\Lambda'' \theta_{a} (\epsilon') \theta_{a}
(\epsilon'') &= \tfrac 12 \Lambda ^{2}_{a} \\ \int d\Lambda' d\Lambda''
\theta_{a} (\epsilon')\theta _{+}(\epsilon'' ) &=2 \Lambda_{a} \Lambda _{b}
- \Lambda ^{2}_{b} \\ \int d\Lambda' d\Lambda'' \theta_{+} (\epsilon')\theta
_{a}(\epsilon'' ) &= \Lambda ^{2}_{b} \\ \int d\Lambda' d\Lambda'' \theta
_{+}(\epsilon' ) \theta _{+}(\epsilon'' ) &=2 \Lambda ^{2}_{b} \end{aligned}
\end{equation}

% \begin{equation} \begin{aligned}
% \int d\Lambda' d\Lambda''  & \begin{pmatrix}
% \theta_{a} (\epsilon')  \theta_{a} (\epsilon'') ,& \theta_{a} (\epsilon')\theta _{+}(\epsilon'' )  \\ 
% \theta_{+} (\epsilon')\theta _{a}(\epsilon'' ),  &   \theta _{+}(\epsilon' )  \theta _{+}(\epsilon'' )  
% \end{pmatrix}    =  \begin{pmatrix}
% \tfrac 12 \Lambda ^{2}_{a}  ,&  2 \Lambda_{a} \Lambda _{b}  -  \Lambda ^{2}_{b}   
% \\   \Lambda ^{2}_{b}  ,  & 2 \Lambda ^{2}_{b}  \end{pmatrix}
% \end{aligned}  %\label{iteratA} \end{equation}

\noindent which leads to some simplification of the second-order corrections
as predicted by the RG equations 
\begin{equation}
\begin{aligned} G_{a}^{\prime \prime } &=  \Lambda ^{2}_{a} \,
(a_{1}^{0} + a_{2}^{0}) \frac {\partial (a_{1}^{0} + a_{2}^{0})}{\partial
G^{0}_{a}} \\ & + ( 2\Lambda_{a} \Lambda _{b} -   \Lambda ^{2}_{b})
a_{3}^{0} \frac {\partial a_{2}^{0}}{\partial G^{0}_{b}} \\ G_{b}^{\prime
\prime } &=  \Lambda ^{2}_{b} \left ( (a_{1}^{0} + a_{2}^{0})\frac
{\partial a_{3}^{0}}{\partial G^{0}_{a}} + a_{3}^{0} \frac {\partial
a_{3}^{0}}{\partial G^{0}_{b}} \right ) \end{aligned}  \label{secondPredict}
\end{equation}

We find that the similar expressions applicable for the first regime $V_{a}%
\agt V_{b}$ are obtained from \eqref{secondPredict} by the replacement $%
\Lambda _{b}\rightarrow \Lambda _{a}$. These results may now be compared
with the explicit calculation of the second order perturbative corrections,
undertaken in the next section.

\subsection{Perturbation theory in second order}

The contribution to the current in the outgoing channel $j$ at position $z$ 
in second order in the interaction may be expressed as

\begin{equation}
\begin{aligned} J_{j}^{(2)}(z) &=-2\int \frac{d\Omega \,d\omega d\omega
^{\prime }}{(2\pi )^{2}}\,\int_{a}^{L}dxdx^{\prime } \\ &\times
\sum_{l,l^{\prime }}\alpha _{l}\alpha _{l^{\prime }}T_{jll^{\prime
}}(x,x^{\prime };\Omega ,\omega ,\omega ^{\prime }) \end{aligned}
\label{current}
\end{equation}

\begin{figure}[h]
\includegraphics[width=0.9\columnwidth]{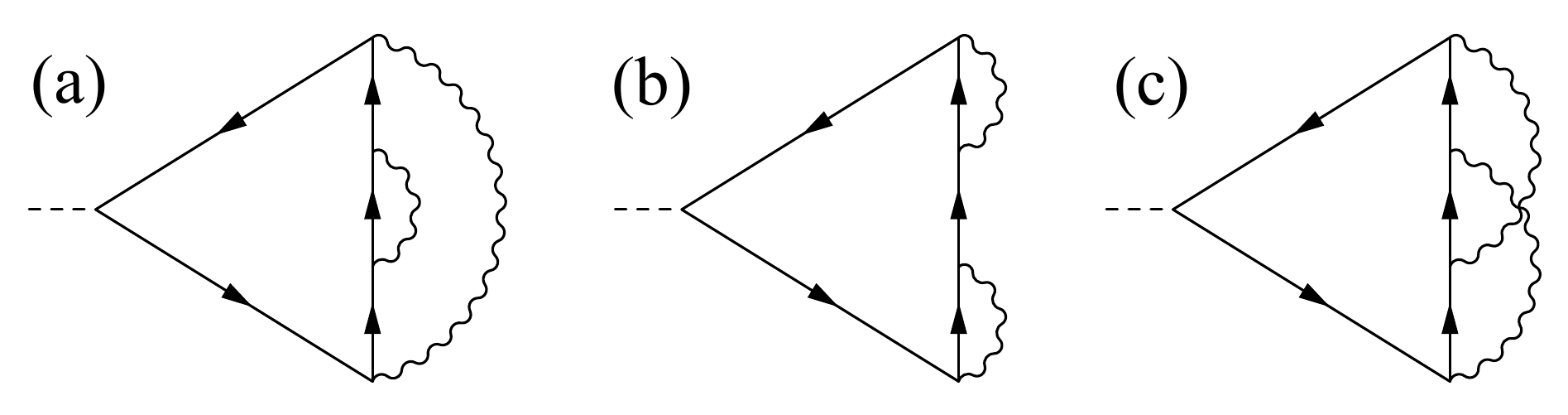}
\caption{Three skeleton diagrams showing the leading contribution to currents in
second order of perturbation}
\label{fig:2order}
\end{figure}

There are three diagrams (see Fig.\ \ref{fig:2order} ) contributing in
second order, each one with arrows both forming a right-handed or a
left-handed loop, which amounts to letting $\Omega \rightarrow -\Omega $,
giving the identical result, thus leading to a prefactor of $2$. The three
diagrams give rise to the combination $T_{jll^{\prime }}(x,x^{\prime
};\Omega ,\omega ,\omega ^{\prime })=T^{(a)}+T^{(b)}+T^{(c)}$\ , where

\begin{equation}
\begin{aligned} T^{(a)} &=\sum_{\mu ,\mu ^{\prime }=1,2}\sum_{\eta _{l},\eta
_{l}^{\prime }=+,-}\mbox{Tr}_{K}[\underline{\gamma
}^{ext}\underline{G}_{\Omega }(j,+,z;l,-\eta _{l},x) \\ &\cdot
\underline{\bar{\gamma}}^{\mu } \underline{G}_{\Omega +\omega }(l,-\eta
_{l},x;l^{\prime },-\eta _{l^{\prime }},x^{\prime })
\underline{\bar{\gamma}}^{\mu ^{\prime }}\\ &\cdot \underline{G}_{\Omega
+\omega +\omega ^{\prime }}(l^{\prime },-\eta _{l^{\prime }},x^{\prime
};l^{\prime },\eta _{l^{\prime }},x^{\prime })\underline{\gamma }^{\mu
^{\prime }} \\ & \cdot \underline{G}_{\Omega +\omega }(l^{\prime },\eta
_{l^{\prime }},x^{\prime };l,\eta _{l},x)\underline{\gamma }^{\mu
}\underline{G}_{\Omega }(l,\eta _{l},x;j,+,z)] \end{aligned} 
\end{equation}
is a rainbow diagram with nested self-energy insertions,  
\begin{equation}
\begin{aligned} T^{(b)} &=\sum_{\mu ,\mu ^{\prime }=1,2}\sum_{\eta _{l},\eta
_{l}^{\prime }=+,-}\mbox{Tr}_{K}[\underline{\gamma
}^{ext}\underline{G}_{\Omega }(j,+,z;l,-\eta
_{l},x)\underline{\bar{\gamma}}^{\mu } \\ &\cdot \underline{G}_{\Omega
+\omega }(l,-\eta _{l},x;l,\eta _{l},x)\underline{\gamma }^{\mu
}\underline{G}_{\Omega }(l,\eta ,x;l^{\prime },-\eta _{l^{\prime
}},x^{\prime })\underline{\bar{\gamma}}^{\mu ^{\prime }} \\ \cdot
&\underline{G}_{\Omega +\omega ^{\prime }}(l^{\prime },-\eta _{l^{\prime
}},x^{\prime };l^{\prime },\eta _{l^{\prime }},x^{\prime })\underline{\gamma
}^{\mu ^{\prime }}\underline{G}_{\Omega }(l^{\prime },\eta _{l^{\prime
}},x^{\prime };j,+,z)] \end{aligned} 
\end{equation}
is a chain diagram with two self-energy insertions in series, and the third
diagram has crossed self-energy insertions

\begin{equation}
\begin{aligned} T^{(c)} &=\sum_{\mu ,\mu ^{\prime }=1,2}\sum_{\eta _{l},\eta
_{l}^{\prime }=+,-}\mbox{Tr}_{K}[\underline{\gamma
}^{ext}\underline{G}_{\Omega }(j,+,z;l,-\eta _{l},x) \\ & \cdot
\underline{\bar{\gamma}}^{\mu } \underline{G}_{\Omega +\omega }(l,-\eta
_{l},x;l^{\prime },-\eta _{l^{\prime }},x^{\prime
})\underline{\bar{\gamma}}^{\mu ^{\prime }}\\ & \cdot \underline{G}_{\Omega
+\omega +\omega ^{\prime }}(l^{\prime },-\eta _{l^{\prime }},x^{\prime
};l,\eta _{l},x) \underline{\gamma }^{\mu } \\ & \cdot \underline{G}_{\Omega
+\omega ^{\prime }}(l,\eta _{l},x;l^{\prime },\eta _{l^{\prime }},x^{\prime
})\underline{\gamma }^{\mu ^{\prime }}\underline{G}_{\Omega }(l^{\prime
},\eta _{l^{\prime }},x^{\prime };j,+,z)] \end{aligned} 
\end{equation}

Here $\underline{G}_{\Omega }$ are $2\times 2$ matrices of Green's functions
in Keldysh space in the absence of interaction, but in the presence of the
scattering effect of the junction, which is expressed in terms of the $S-$%
matrix elements $S_{ij}$, as presented above. The dependence on the
coordinates may be split off:

\begin{equation}
T_{jll^{\prime }}^{(a,b,c)}(x,x^{\prime };\Omega ,\omega ,\omega ^{\prime
})=e^{-2i(\omega x+\omega ^{\prime }x^{\prime })}T_{jll^{\prime
}}^{(a,b,c)}(\Omega ,\omega ,\omega ^{\prime })
\end{equation}

The trace $\mbox{Tr}_{K}$ is over the lower (fermionic) Keldysh indices; the
fermion-boson vertices, $\gamma _{ij}^{\mu }\rightarrow \underline{\gamma }%
^{\mu }$, $\bar{\gamma}_{ij}^{\mu }\rightarrow \underline{\bar{\gamma}}^{\mu
}$, tensors of rank $3$\ defined in Keldysh space, are given by

\begin{equation}
\gamma _{ij}^{1}=\bar{\gamma}_{ij}^{2}=\tfrac{1}{\sqrt{2}}\delta _{ij},\quad
\gamma _{ij}^{2}=\bar{\gamma}_{ij}^{1}=\tfrac{1}{\sqrt{2}}\tau _{ij}^{1},
\label{def:gamma}
\end{equation}%
with $\tau ^{1}$ the first Pauli matrix. The external vertex is given by

\begin{equation}
\gamma _{ij}^{ext}=\frac{i}{2}\left( 
\begin{array}{cc}
1 & 1 \\ 
-1 & -1%
\end{array}%
\right) =\frac{i}{2}\left( 
\begin{array}{c}
1 \\ 
-1%
\end{array}%
\right) \left( 
\begin{array}{cc}
1 & 1%
\end{array}%
\right) ,
\end{equation}%
which suggests to interpret the trace in Keldysh space as operating with the
Keldysh matrices on the vector $\left( 
\begin{array}{cc}
1 & -1%
\end{array}%
\right) ^{T}$ and forming the inner product of the resulting vector with the
vector $\frac{i}{2}\left( 
\begin{array}{cc}
1 & 1%
\end{array}%
\right) .$

The calculation of second order corrections to the currents $J_{a}$, $J_{b}$
is a tedious procedure and is discussed in more detail in the Appendix. Here
we provide the summary of this calculation. We find the corrections in the
form

\begin{equation}
\begin{aligned} G ''_{a} &= B_{a1} F_{a1} + B_{a2} F_{a2} +
B_{a3} F_{a3} , \\ G''_{b} &=  B_{b1}F_{b1} + B_{b2}F_{b2} +
B_{b3} F_{b3}  \end{aligned} 
\end{equation}
with

\begin{equation}
\begin{aligned} B_{a1} & = \frac{1}{16} G_b \left( G_a-1 + g_{3} \left(2
G_a-1\right)\right) \\ & \times \left( 4 G_a+3 G_b-4 +g_{3}\left(6
G_b-4\right)\right), \\ B_{a2} & =-  \left(2 G_a-1\right) \left(4
G_a^2-4 G_a+G_b\right), \\ B_{a3} & = -\frac{1}{8} \left(1+2 g_{3}\right)
G_b \left(4 G_a^2-4 G_a+G_b\right), \\ B_{b1} & = \frac{1}{16} G_b \left( 2
G_a+G_b-2 +2 g_{3}\left(G_b-1\right)\right) \\ & \times \left( 4 G_a+3 G_b-4
+g_{3}\left(6 G_b-4\right)\right), \\ B_{b2} & = \frac{1}{4} G_b \left(
\left(1+2 g_{3}\right) G_a G_b-4(G_a-G_{a}^{2}) -g_{3}G_b\right), \\ B_{b3}
& = \frac{1}{16} G_b^2 \left( {G_b} +4 g_{3}(1+g_{3}) \left(G_b-1\right)
\right) . \end{aligned} 
\end{equation}

Coefficients $F_{j}$ are defined as integrals over energy, they are
independent of $G_{a,b}$ and are discussed in the Appendix.

We distinguish again between two regimes: i) $V_{a}\agt V_{b}$ and ii) $%
V_{a}\ll V_{b}$. In second regime we find, using the results given in the
Appendix 
\begin{equation}
\begin{aligned} F_{a1} = & \Lambda_{b}^{2}, \quad F_{b1} = 2
\Lambda_{b}^{2}, \\ F_{a2} = & - \Lambda_{a}^{2} , \quad F_{b2} = 2
\Lambda_{b}^{2}, \\ F_{a3} = & \Lambda_{b}^{2} - 4 \Lambda_{a} \Lambda_{b} ,
\quad F_{b3} = 2 \Lambda_{b}^{2}. \end{aligned}  \label{Lambdalow}
\end{equation}%
which gives 
\begin{equation}
\begin{aligned} G''_{a} &= B_{a1} \Lambda^{2}_{b}- B_{a2} \Lambda_{a}^{2}+
B_{a3}( \Lambda_{b}^{2} - 4 \Lambda_{a} \Lambda_{b}) , \\ G''_{b} &= 2
(B_{b1} + B_{b2} + B_{b3} ) \Lambda^{2}_{b}, \end{aligned}
\label{correcExplicit}
\end{equation}%
in the first regime we should merely replace $\Lambda _{b}\simeq \Lambda _{a}
$ in these expressions. In both cases one can check the equivalence of %
\eqref{secondPredict} and \eqref{correcExplicit}, which proves that the
second order corrections are indeed exactly generated by the RG equations %
\eqref{Beta}.

Alternatively, we checked the validity of \eqref{Beta} by application of
Eq.\ \eqref{dG/dLambda}. In the non-trivial second regime $V_{a}\ll V_{b}$,
we use the above expressions, \eqref{correcExplicit} and find that the terms 
$\sim \alpha ^{2}$ in \eqref{dG/dLambda} are proportional to $(\Lambda
_{a}-\Lambda _{b})\,d\Lambda _{b}/d\Lambda $, which is identically zero.

\subsection{RG-equations to second order}  

\begin{figure}[h]
\includegraphics[width=0.5\columnwidth]{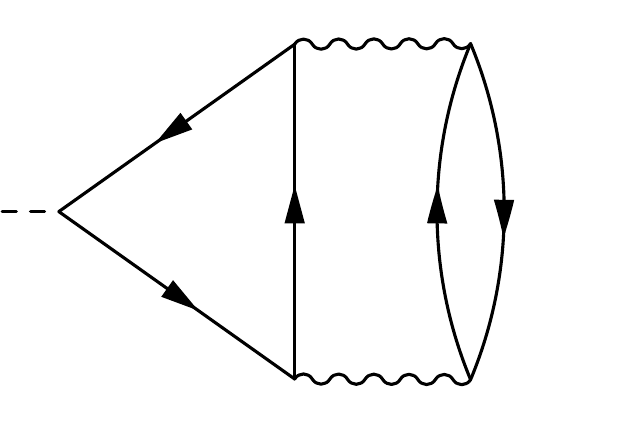}
\caption{A skeleton diagram showing the subleading contribution to currents in
second order of perturbation}
\label{fig:subl}
\end{figure}

As shown previously, \cite{Aristov2014} 
there are also contributions linear in $\Lambda$ to the conductances in second order of the interaction , i.e. subleading terms $\sim \alpha^{2} \Lambda$. 
These contributions arise from the diagram shown in Fig.\ \ref{fig:subl}, featuring two fermion loops.
Terms linear in $ \Lambda$ generate contributions to the RG beta functions. They give rise to
$\alpha^{2}$-corrections to the scaling exponents.  The modification of the subleading terms in multi-lead junctions out of equilibrium was not previously analyzed. 

Performing calculations similar to the one described in the Appendix we arrive at the following results. 
The beta functions in Eq. \eqref{Beta}  retain their general structure, 
with updated coefficient functions $a_j$ . We have to replace 
\begin{equation}
\begin{aligned} a_{1} & 
\to  (1 - \alpha (G_{a}-\tfrac 12)) \, a_{1}  ,   \\
a_{2}  & \to a_{2} + 
\tfrac\alpha{8} G_{b}  (G_{a}-\tfrac 12) \\
& \times (1-\tfrac14 G_{b} 
- g_{3} (G_{b}+g_{3}(G_{b}-1)) 
) ,  \\
a_{3} & \to a_{3}   
-\tfrac{\alpha}{4} G_{b} \Big (1 + 2G_{a}  (G_{a}-1) - \tfrac 34 G_{b} 
\\ & + \tfrac 14 
( G_{b} +2g_{3}(G_{b}-1))^{2} \Big )
. \end{aligned} 
\label {a123mod}
\end{equation}
For the detached third wire, $G_{b}= 0$, we obtain $a_{2} (G_{a},0) =a_{3} (G_{a},0) =0 $ and the
modification of $a_{1} (G_{a},0)$ is in accordance with the second order expansion of Eq.\ (47) in  Ref. \cite{Aristov2014}. 
Notice that for the pure tunneling case, \cite{Aristov2010} when $G_{b} = 4G_{a} (1-G_{a})$, the part with $\theta _{a }(\epsilon )$ disappears, $a_{1}=0$, and the two RG equations with $\theta_{+}(\epsilon)$ become linearly dependent, since  in this case  $\tfrac {d}{d\Lambda}(G_{b} - 4G_{a} (1-G_{a}))=0$. \cite{Aristov2017a}  

At the same time there is not much simplification in \eqref{a123mod} in the case of $V_{a} \sim V_{b}$, when we can let  $ \theta_{+}(\epsilon) \simeq  2\theta_{a}(\epsilon)$. The remaining expressions are complicated, as can be seen, e.g. by expanding Eq.\ (12) in \cite{Aristov2013} in powers of $\alpha$ at $c=0$.

\section{Summary}

In this paper we established the validity of the RG equations for the
conductances of multilead junctions of Tomonaga-Luttinger liquid wires in a
situation out of equilibrium. Comparing to the equilibrium case, when the RG
flow stops at some unique cutoff, which characterizes the low-energy scale
of the whole system of wires, the out-of-equilibrium situation can be
characterized by several such scales, referring to $(N-1)$ relative voltages
between the $N$ wires. In this situation it is not clear which of these
scales should be used as a cutoff in the corresponding expressions.
Previously we found \cite{Aristov2017a} that the RG equations contained
several functions, describing partial stops of the RG flow, so that the
direction of the flow could alter during the renormalization process. In
this paper we formulate the statement about the scaling property for the set
of conductances, characterizing a general setup with $N$ wires. Then we
consider the particular example of the Y junction ($N=3$) with different
strength of interaction in the main wire and the tunneling tip. We focus on
the two most interesting regimes, when i) all voltages are of the same order
and ii) the voltage $V_{a}$ in the main wire is much smaller than the
voltage $V_{b}$ at the tip.

The second order corrections are calculated in two ways. One way is the
iteration of the RG equations to second order, which is less trivial in the
presence of several cutoffs. A second way is the direct calculation of
second order corrections by means of computer algebra, which requires
considering a large number of partly canceling contributions. We find that
both ways of calculation lead to identical results in both regimes. 
As a by-product we derived the corrections to the beta functions of second order in the interaction.

We believe that our results may be useful for a generalization of ideas of
scaling in the presence of several low energy cutoffs, appearing
particularly in out-of-equilibrium situations.

\section{Acknowledgements}

The work of D.A. was partly supported by RFBR grant No. 15-52-06009 .

\appendix

\section{Details of calculation}

The expression for the corrections \eqref{current} requires five
integrations.

Let us first discuss the integration over $x,x^{\prime }$. The dependence of
each Green's function in \eqref{def:Green} on the coordinates comes from two
factors: the step functions, $\theta (x)$, and the oscillatory exponentials, 
$e^{i\omega x}$. The outgoing current is determined at a point $z$ in the
lead, which is outside the interacting region. In our terms this means that
the coordinate $z$ is greater that any other of the coordinates, $x$, $%
x^{\prime }$. This allows to simplify the step functions by replacing $%
\theta (z-x)=1$, $\theta (x-z)=0$, etc. The corrections to the incoming
currents are zero, which is verified by putting $\theta (z-x)=0$, $\theta
(x-z)=1$. The exponents $e^{\pm i\omega x}$ do not contain $\Omega $, and
after appropriate change of sign in $\omega $ , $\omega ^{\prime }$ can be
reduced to unique form $e^{-2i(\omega x+\omega ^{\prime }x^{\prime })}$. The
remaining expressions may still contain $\theta (x-x^{\prime })$, $\theta
(x^{\prime }-x)$, however, after symmetrization, $x\leftrightarrow x^{\prime
}$, $\omega \leftrightarrow \omega ^{\prime }$, these stepwise functions
combine to unity.

The integration over $x$, $x^{\prime }$ is now simple, since the dependence
on the coordinates in each term is reduced to $e^{-2i(\omega x+\omega
^{\prime }x^{\prime })}$. We have 
\begin{equation*}
\int_{a}^{L}dx_{1}e^{-2i\omega x}=\frac{e^{-2i\omega a}-e^{-2i\omega L}}{%
2i\omega }\rightarrow \frac{1}{2i\omega }
\end{equation*}%
where the last equality is obtained because the rapidly oscillating factor $%
e^{-2i\omega L}$ is only important as an infrared cutoff at the smallest $%
\omega $, and in our case this cutoff is provided by the voltages. The
integration over $x$,$x^{\prime }$ hence leads to the overall factor, $%
-1/(4\omega _{1}\omega _{2})$.

It is convenient to symmetrize the appearing expressions with respect to $%
\omega \rightarrow -\omega $, picking the odd-in-$\omega $ part of the
integrand, and then to consider a positive interval of energies in
subsequent integrations : 
\begin{equation}
\int_{0}^{\omega _{0}}\frac{d\omega \,d\omega ^{\prime }}{4\,\omega \omega
^{\prime }}  \label{symint1}
\end{equation}%
with $\omega _{0}$ ultraviolet cutoff.

Let us now discuss the integration over $\Omega$. In general, we find terms,
linear in $h_{l}(\omega)=\tanh [(\omega -\mu _{l})/2T]\equiv h_{0} (\omega
-\mu _{l})$, and cubic in this quantity, $\sim h_{l} h_{m} h_{j}$. The
quadratic terms, $h_{l}h_{m}$, disappear.

Every cubic combination has the form $h_{0}(\Omega _{1})h_{0}(\Omega
_{2})h_{0}(\Omega _{3})$, with $\Omega _{j}=\Omega +\ldots $, (e.g. $\Omega
_{j}=\Omega -\mu _{1}+\omega $ or $\Omega _{j}=\Omega -\mu _{2}+\omega
-\omega ^{\prime }$). In order to regularize the integral over $\Omega $, we
subtract a term $h_{0}(\Omega _{3})$ so that the combination $(h_{0}(\Omega
_{1})h_{0}(\Omega _{2})-1)h_{0}(\Omega _{3})$ is convergent at $\Omega
\rightarrow \pm \infty $. All regularization terms $h_{0}(\Omega _{3})$ are
combined with the other terms of the first power in $h_{0}(\ldots )$.

In the so regularized terms we may shift the argument and write 
\begin{equation}
\begin{aligned} &\int d\Omega (h_{0}(\Omega+\tilde\omega_{1})
h_{0}(\Omega+\tilde\omega_{2}) -1 ) h_{0}(\Omega+\tilde\omega_{3}) , \\ & =
\int d\Omega (h_{0}(\Omega+\tilde\omega_{1}-\tilde\omega_{3})
h_{0}(\Omega+\tilde\omega_{2}-\tilde\omega_{3}) -1 ) h_{0}(\Omega) , \\ &
\equiv f_{3}
(\tilde\omega_{1}-\tilde\omega_{3},\tilde\omega_{2}-\tilde\omega_{3})
\end{aligned}  \label{def:f3}
\end{equation}%
with 
\begin{equation}
\begin{aligned} f_{3} (A,B) & = \int d\Omega \, \tanh \tfrac\Omega{2T} (
\tanh \tfrac{\Omega+A}{2T} \tanh\tfrac{\Omega+B } {2T} -1 ) , \\ & = 2 \coth
\tfrac{A-B}{2T} \left( A \coth\tfrac{A}{2T} - B \coth \tfrac{B}{2T} \right)
, \\ & \to 2 (|A| - |B|)\mbox{ sign}|A-B| \,, \end{aligned}  \label{f3expl}
\end{equation}%
the last line obtained in the limit $T\rightarrow 0$, which we are mostly
interested in.

The terms linear in $h_{0}(\ldots)$ cancel each other. This can be proved in
two steps. First we subtract one and the same term $h_{0}(\Omega)$ from each
term, $h_{0}( \Omega+\tilde\omega_{1}) \to h_{0}( \Omega+\tilde\omega_{1}) -
h_{0}(\Omega)$, making the integral convergent: 
\begin{equation}
\int d\Omega (h_{0}(\Omega+\tilde\omega_{1}) - h_{0}(\Omega) ) = - 2
\tilde\omega_{1} \,.  \label{def:f1}
\end{equation}
The combination of all such terms does not contain $\omega$, $\omega^{\prime
}$ and hence vanishes when performing the symmetrization $\omega \to -\omega$%
, leading to above expression \eqref{symint1}. As a second step we sum all
terms with $h_{0}(\Omega)$ and verify that they cancel each other.

Let us discuss further simplifications. When obtaining the terms, $%
f_{3}(\ldots)$, the third argument $\Omega+\tilde\omega_{3} $ in Eq.\ %
\eqref{def:f3} was chosen by computer, i.e. almost randomly from the human
viewpoint. It means that many appearing terms may look differently but lead
to the same result after subsequent integrations. To get rid of this
ambiguity we use the symmetry properties 
\begin{equation}
\begin{aligned} \mbox{(i) } f_{3}(a,b) & = f_{3}(b,a) , \quad \mbox{(ii) }
f_{3}(a,b)= - f_{3}(-a,-b) , \\ \mbox{(iii) } f_{3}(a,b) & \approx
f_{3}(a-b,-b) , \end{aligned}  \label{symf3}
\end{equation}
where the last approximate equality means generation of the linear in $h_{0}$
terms which eventually disappear. The last equivalence property is (iv) the
symmetry with respect to $\omega \leftrightarrow \omega^{\prime }$.

We find that the number of expressions to be considered is strongly reduced,
when we take each correction term of the form $Af_{3}(a,b)$, strip its
prefactor $A$ and perform the symmetry operations, (i) -- (iv), for $%
f_{3}(a,b)$. We thus form an equivalence list of length $2^{4}=16$ which is
ordered according to the computer's internal rules. We choose a first
element $f_{3}(a_{\ast },b_{\ast })$ of this ordered list as a
representative. Such operation leaves only the factors $f_{3}(a_{\ast
},b_{\ast })$ which are not related by symmetry operations, i.e. are
essentially different.

Note that at this step of our analysis we may find terms of the form $%
f_{3}(\omega ^{\prime }-\omega +a,b)$ with $a,b$ dependent only on the $\mu
_{j}$. Recalling the initial expression $e^{-2i(\omega x+\omega ^{\prime
}x^{\prime })}$ we see that a shift $\omega ^{\prime }=\bar{\omega}+\omega $
and integration over $\omega $ leads to $\delta (x+x^{\prime })$. In
combination with the condition $x,x^{\prime }>0$ this means that such terms
should be discarded.

In the intermediate expressions for the three types of diagrams in Fig.\ \ref%
{fig:2order} we found terms containing neither $\mu _{1}$ nor $\mu _{2}$.
Such unphysical terms cancel in the combination of all three types of
diagrams.

To condense the expressions further we introduce the symmetric combination 
\begin{equation}
\begin{aligned} f_{s}( A , B ) & = -f_3\left(-\omega  - A,\omega'
-B\right)+f_3\left(-\omega + A ,\omega' +B \right) \\ & +f_3\left(\omega
-A,\omega' -B\right)-f_3\left(\omega +A, \omega ' + B\right) \end{aligned}
\label{def:fsym}
\end{equation}%
with the properties 
\begin{equation}
\begin{aligned} f_{s} ( A, B) & \simeq 8 A \,,\quad \omega \ll \omega' , \\
& \simeq 8 B \,,\quad \omega \gg \omega' , \\ f_{s} ( A, B) |_{\omega=0}& =
f_{s} ( A, B) |_{\omega'=0}= 0 \end{aligned}  \label{fs2}
\end{equation}%
From the piecewise linear form of \eqref{f3expl} it is clear that the
transition between the values $f_{s}\simeq 8A$ and $f_{s}\simeq 8B$ takes
place at $|\omega -\omega ^{\prime }|\alt|A-B|$.

The corrections to the currents are then expressed through integrals

\begin{equation}
\begin{aligned} \{ & F_{a1}, F_{a2} , F_{a3} ,F_{b1} ,F_{b2} , F_{b3} \} \\
= & -\int_{0}^{\omega_{0}} \frac {d \omega d \omega'  }{4\omega 
\omega ' } \Big \{f_{s}\left( \mu_1,0\right) - f_{s}\left( \mu_2, 0\right)
, \\ & f_{s}\left( \mu_2-\mu_1, 0\right), \\ & f_{s}\left(\mu_2-\mu_1,
\mu_2\right) +f_{s}\left( \mu_2-\mu_1, -\mu_1\right) , \\ & f_{s}\left(
\mu_1,0\right) + f_{s}\left( \mu_2, 0\right) , f_{s}\left( \mu_1,
\mu_2\right), \\ & f_{s}\left(\mu_2-\mu_1, \mu_2\right) - f_{s}\left(
\mu_2-\mu_1, -\mu_1\right) \} \end{aligned}  \label{defFab}
\end{equation}%
We have here a generic integral 
\begin{equation}
\begin{aligned} & \int_{0}^{\omega_{0}} \frac {d \omega d \omega' }{4\omega
\omega ' } f_{s} (A, B) \\ & = A \ln^{2} \frac{\omega_{0}}{|A|} + B\ln^{2}
\frac{\omega_{0}}{|B|} \, , \quad |A| \sim |B| , \\ 
% & = A \ln^{2} \frac{\omega_{0}}{|A|} + B \left (\ln^{2} \frac{\omega_{0}}{|A|} + 2
% \ln\frac{\omega_{0}}{|A|} \ln \frac{|A|}{|B|} \right ) , \\ 
& = (A-B)
\ln^{2} \frac{\omega_{0}}{|A|} +2 B\ln \frac{\omega_{0}}{|A|} \ln
\frac{\omega_{0}}{|B|} \, , \quad |A| \gg |B| , \end{aligned}
\label{estimate}
\end{equation}%
we use these estimates for the calculation of the results \eqref{Lambdalow}
in the main text.

%%%%%%%%%%%%%%%%%%%%%%%%%%%%%%%%%%%%
%
%%%%%%%%%%%%%%%%%%%%%%%%%%%%%%%%%%%%

%\bibliography{../../../../Bibliography/MainBib.bib,../../../../Bibliography/onedimen.bib,../../../../Bibliography/oldies.bib}    \end{document}

\end{document}